\documentclass[journal]{IEEEtran}
\pdfoutput=1
\usepackage{nopageno}
\usepackage[pdftex]{graphicx}
\usepackage{amsmath,amsfonts,amssymb}

\DeclareMathOperator*{\argmin}{arg\,min}

\ifCLASSINFOpdf
\else
\fi
\begin{document}
%
\title{A directional total variation minimization algorithm for isotropic
resolution in\\ digital breast tomosynthesis}
%
%
%

\author{Emil~Y.~Sidky, Xiangyi~Wu, Xiaoyu~Duan, Hailing~Huang,
        Wei~Zhao, Leo~Y.~Zhang, John~Paul~Phillips, Zheng~Zhang, Buxin~Chen,
        Dan~Xia, Ingrid~S.~Reiser, Xiaochuan~Pan
\thanks{E. Y. Sidky, J. P. Phillips, Z. Zhang, B. Chen, D. Xia, I. S. Reiser,
and X. Pan are with the Department of Radiology MC-2026,
The University of Chicago,
5841. S. Maryland Ave., Chicago, IL, 60637.}%
\thanks{X. Wu, X. Duan, H. Huang, and W. Zhao
are with the Department of Radiology, Renaissance School of Medicine,
Stony Brook University,
101 Nicolls Road,
Health Sciences Center, Level 4,
Stony Brook, NY, 11794.}%
\thanks{L. Y. Zhang is with the Department of Mathematics,
University of Washington, Box 354350, C-138 Padelford,
Seattle, WA 98195-4350.}%
}

\maketitle
\thispagestyle{empty}

\begin{abstract}
An optimization-based image reconstruction
algorithm is developed
for contrast enhanced digital breast tomosynthesis (DBT) using dual-energy scanning.
The algorithm minimizes directional
total variation (TV) with a data discrepancy and non-negativity constraints.
Iodinated contrast agent (ICA) imaging is performed by
reconstructing images from dual-energy DBT data followed
by weighted subtraction.
Physical DBT data is acquired with a Siemens Mammomat
scanner of a structured breast phantom with ICA inserts.
Results are shown for both directional TV minimization and
filtered back-projection for reference.
It is seen that directional TV is able to substantially reduce depth blur for the ICA
objects.
\end{abstract}


%
\IEEEpeerreviewmaketitle

\section{Introduction}
\label{sec:introduction}

Over the past two decades, X-ray based tomosynthesis has been an emerging imaging modality
that has been making great strides in clinical adoption; most notably, digital breast tomosynthesis (DBT)
is becoming the workhorse imaging device for breast cancer screening. Although DBT yields partially
tomographic volumes, there is still a large gap between it and X-ray computed tomography in terms
of achieving isotropically high resolution and in the ability to achieve quantitive imaging
for shape and volume features of tissues within the scanned subject. Tomosynthesis imaging
is subject to significant depth blur, and this blurring is proportional to the size of
the imaged structures. Because of this depth blurring it is not possible to outline the boundaries
of structures and estimate their volume. Moreover, the gray scale values in the image do not
directly correspond to a physical density.

In this work, an optimization-based algorithm is developed that may enable quantitive imaging
with limited angular range scanning such as what is performed in a tomosynthesis acquisition.
In order to accomplish this, we exploit sparsity in the gradient magnitude of the scanned subject.
For sparse-view imaging over a full scanning angular range, the use of total variation (TV) minimization
has proven effective \cite{sidky2008image}. For limited angular range scanning, however, directional
TV constraints show promise \cite{zhang2021directional}. The use of directional TV constraints can be shown
to be equivalent to directional TV minimization where the gradient magnitudes in different directions
have different penalty weights \cite{leo2023}. The work presented in Ref. \cite{leo2023} showed only simulation
studies, and in this work we apply this algorithm to physical DBT transmission data.

The application of interest is contrast enhanced (CE) DBT \cite{huang2019comparison},
where the goal is to obtain a quantitative image of the Iodinated contrast agent (ICA) distribution.
If such a system is successfully developed it could enable the use of CE-DBT in assessment of breast cancer
therapy effectiveness. For this preliminary
work, DBT data is obtained for structured breast phantom with ICA inserts.
Volumes are reconstructed from scans at 30 and 49 kV, and the weighted image subtraction method is used
to isolate the ICA distribution. 

\section{Methods}
\label{sec:methods}

\subsection{Image reconstruction}

In order to obtain reconstructed DBT volumes that have nearly isotropic resolution
for the ICA distribution, we exploit the fact that this distribution is localized
with a fairly simple structure and high contrast. In the algorithm
implementation directional gradient sparsity regularization is employed and the
resolution is reduced by a factor of eight from the native resolution of the DBT
system's detector. 

The image reconstruction algorithm is developed for the Siemens Mammomat scanner
that acquires 25 projections over a 50 degree scanning arc and the
detector resolution is 85 microns.
The transmission data are processed with the negative logarithm after dividing by a flood-field
scan and the relationship between this processed data and the DBT volume is assumed
to be standard linear X-ray projection as described by
\begin{equation*}
g = X f,
\end{equation*}
where $f$ is a discrete representation of the DBT volume, $X$ is a matrix
encoding X-ray projection, and $g$ is the processed projection data.

For sparsity regularization, image reconstruction is formulated using an
optimization model where weighted image directional total variation (TV) is minimized
subject to a data-fidelity constraint. Furthermore, an $\ell_1$ penalty term
is included that encourages pixel sparsity and helps to confine the image to the
true object support. Finally, the data fidelity term compares ramp-filtered data
with ramp-filtered estimated data, which has two important effects: this filtering
deemphasizes low spatial-frequency discrepancy in the sinogram estimation and
it also performs preconditioning.
\begin{align*}
f = \argmin_{f^\prime} \bigl\{ \alpha \| \nabla_x f^\prime \|_1 +
\alpha \| \nabla_y f^\prime \|_1 + (2-\alpha) \| \nabla_z f^\prime \| \bigr. \\
\bigl. + \beta  \|f^\prime \|_1
\text{  such that  } \|R( g - X f^\prime) \|_2 \le \epsilon \text{ and } f^\prime \ge 0 \bigr\}.
\end{align*} 
The $x$ and $y$ coordinates are in-plane with $y$ being along the direction of travel
for the X-ray source, and the $z$ direction is the depth direction; i.e. perpendicular
to the X-ray detector. The symbol $\nabla$ represents finite differencing along the direction
indicated by the subscript; $\alpha$ is a weighting parameter that should be chosen
between 0 and 2; $\beta$ is the weighting parameter for the $\ell_1$ penalty,
$R$ represents ramp filtering and $\epsilon$ is a data error tolerance parameter. 
The posed optimization problem is convex and it can be solved with the primal-dual
algorithm of Chambolle and Pock \cite{chambolle2011first}.

\begin{figure}[!t]
\centering
\includegraphics[width=\columnwidth]{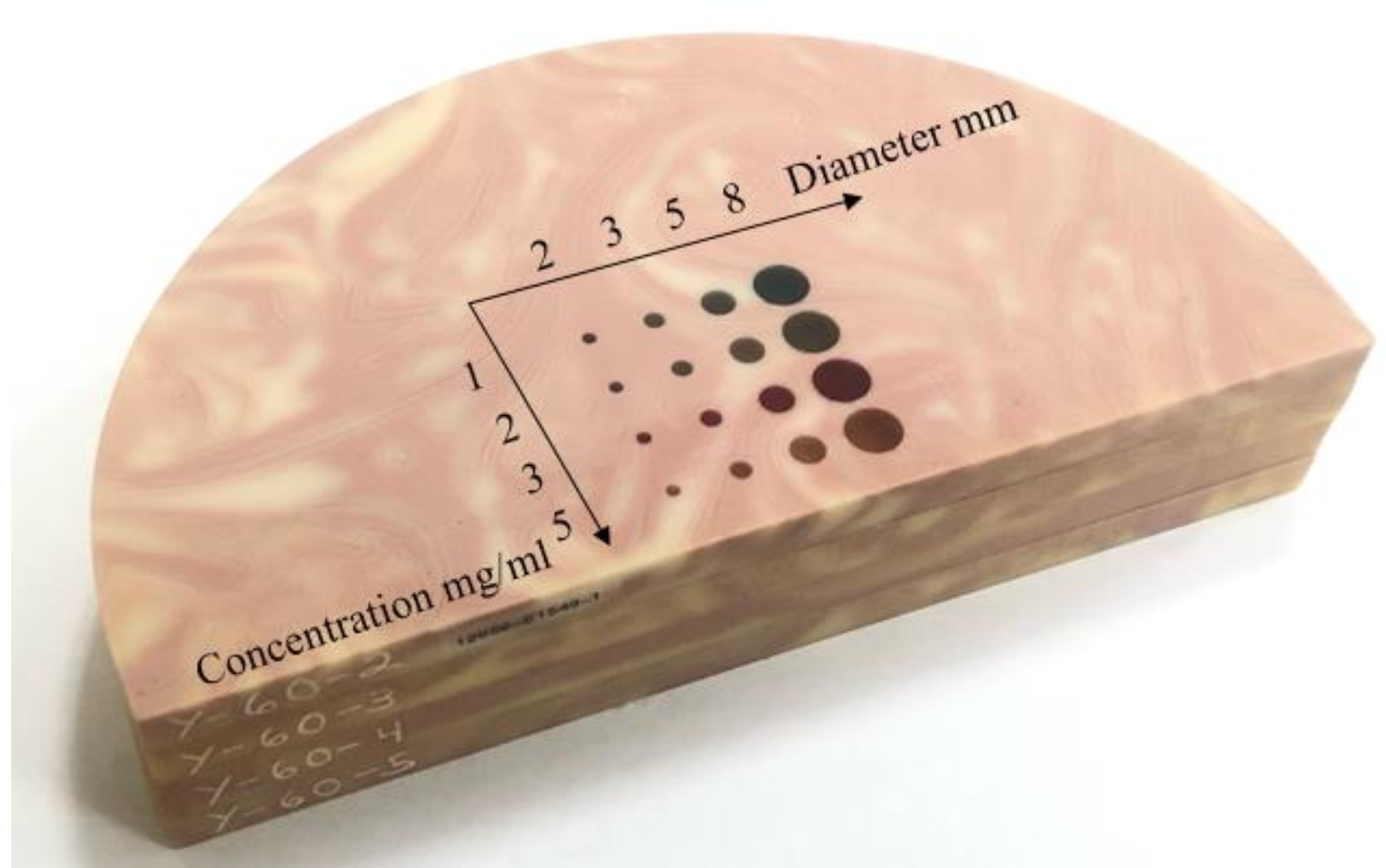}
\caption{CIRS phantom with solid ICA inserts. The background is hetergeneous with 50\% glandularity.}
\label{fig:phantom}
\end{figure}

When $\alpha=1$, the gradient sparsity regularization term is equivalent to TV
and it has been shown that constrained TV minimization is effective for accurate
image reconstruction for sparse-view acquisition when the scanning arc is complete.
Recently, we have shown that the anisotropic weighting of the directional TV terms
can improve image reconstruction accuracy for limited scanning angular ranges such
as those used for DBT. This in combination with use of a coarse volume grid allows
for nearly isotropic resolution in the reconstructed DBT images using directional TV
minimization.

\subsection{Image post-processing}

Even when the transmission data preprocessing accurately estimates the image
sinogram, Reconstructed DBT images are susceptible to low spatial-frequency artifact
due to the limited angular range of the scan. In order to facilitate the dual-energy image
subtraction the low spatial-frequency artifacts are estimated and removed by normalization.
Specifically, the low spatial-frequency image component is estimated by robust polynomial
fitting
\begin{equation}
\label{polyfit}
b = \argmin_c \bigl\| m \cdot (s - Pc) \bigr\|_1,
\end{equation}
where $s$ is an in-plane slice of the image $f$; $P$ is a matrix of low-order polynomials; $c$
is a vector of coefficients; $m$ is an indicator function which is one on the object support
and zero otherwise, where the object support is estimated by thresholding the image slice, $s$;
and $b$ is the estimated background for the slice $s$. The use of the $\ell_1$-norm for the
objective function encourages the difference image in its argument to be sparse, and this design
should preserve important small structures in the image when the background drift is processed away.

\begin{figure}[!t]
\centering
\includegraphics[width=\columnwidth]{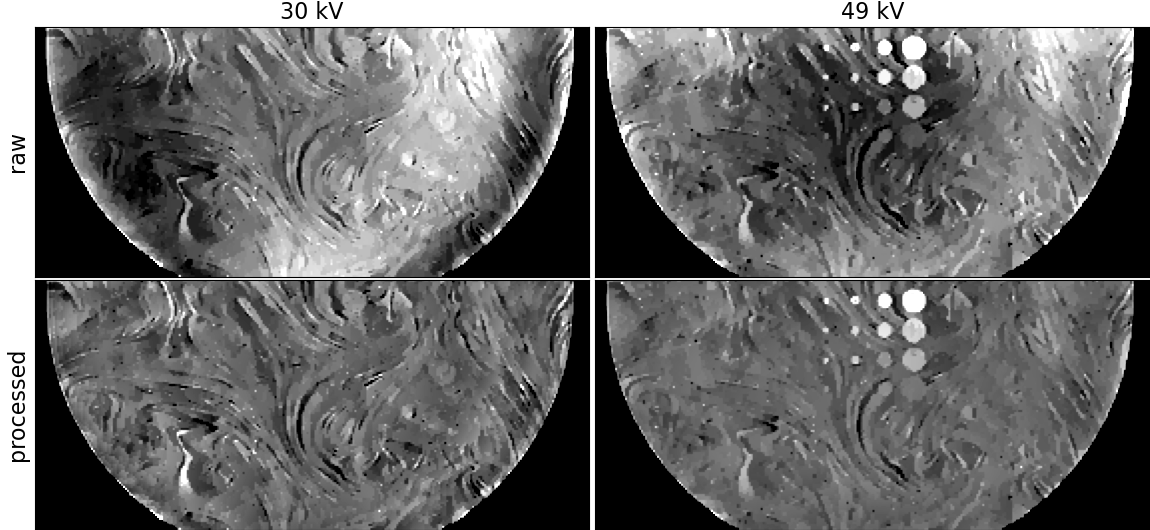}
\caption{Raw and processed in-plane images containing all ICA objects for the directional TV minimization 
images for the 30 kV (Left) and 49 kV (Right) DBT scans. Image processing is performed by polynomial
fitting and normalization described in Eqs. (\ref{polyfit}) and (\ref{improc}), respectively. The gray scale
for the raw images is selected to span the range of background variation, and the gray scale for the 
normalized images is selected to be $\pm$ 30\% of background.
}
\label{fig:imageProc}
\end{figure}

The polynomial fitting is performed
slice by slice with the result stacked into the background image $f_b$. Likewise, the mask image
$f_m$ is assembled by stacking the slice masks. The algorithm for solving this robust polynomial
fitting is iteratively reweighted least-squares.
The final displayed image, $f_d$, is computed as follows
\begin{equation}
\label{improc}
f_d = f_m \cdot \frac{f}{f_b}.
\end{equation}
After the processed images are computed for low and high kV, a difference image is computed
\begin{equation*}
f^{(ICA)} = \left( f^{(HE)}_d - 1 \right)  - w \cdot \left( f^{(LE)}_d -1 \right).
\end{equation*}
For this work, the dual energy data is processed in the image domain by performing
a weighted subtraction of low and high kV reconstructed volumes, where the weights
are designed to highlight the ICA distribution. Because the low spatial-frequency
background variation is reduced by normalization, the background of the processed
images center on the value of 1; thus 1 is subtracted from both images before
performing the weighted subtraction. The weighting parameter $w$ is selected by visual inspection.

\begin{figure*}[!t]
\centering
\includegraphics[width=0.9\linewidth]{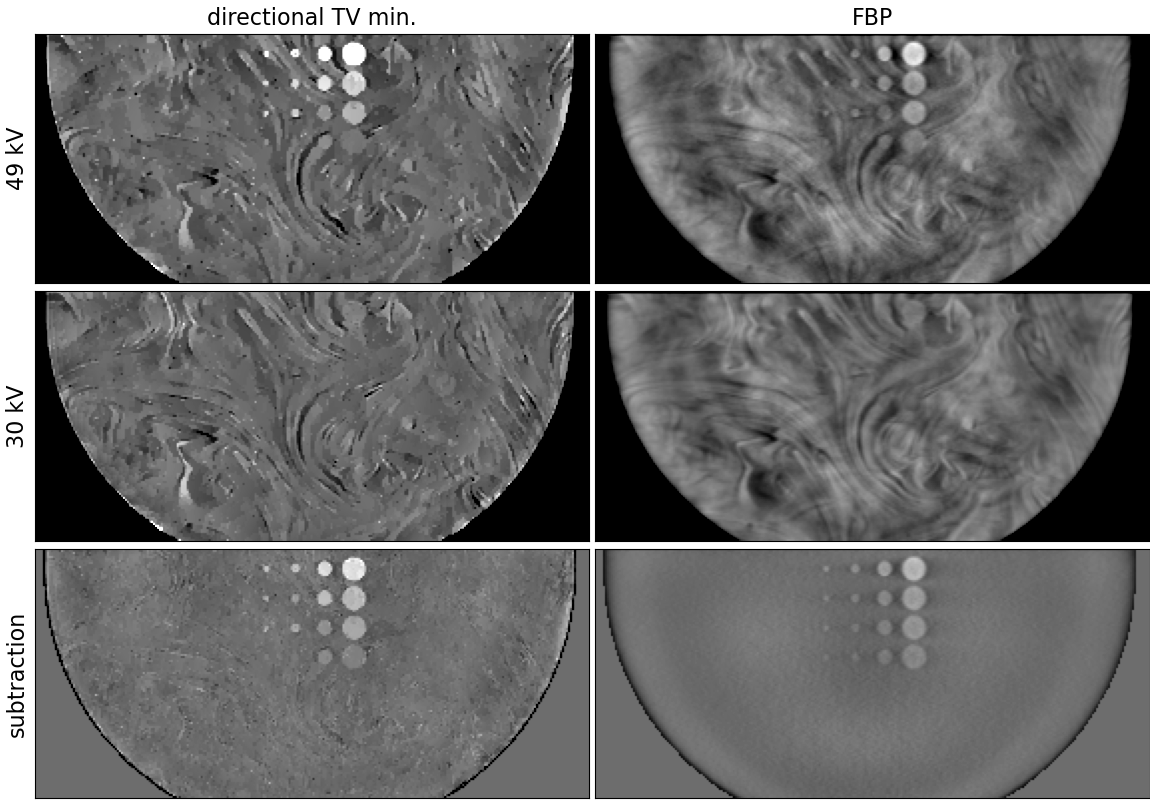}
\caption{In-plane images containing all ICA objects for directional TV minimization (Left) and FBP (Right).
The top and middle rows show the images corresponding to the 49 and 30 kV acquisitions, respectively.
The bottom row shows the weighted subtraction designed to isolate the ICA distributions. The gray scales
on the left and right are matched and the images are masked to the object support.}
\label{fig:inplane}
\end{figure*}

\subsection{DE-DBT transmission data acquisition }
The DE-DBT scanning consists of two DBT acquisitions taken at source potential of 30 and 49 kV using
filtration with Al and Ti, respectively.
DE-DBT scans of a 4 cm thick BR3D phantom (Model 020, CIRS Inc. Norfolk, VA) with solid ICA inserts,
see Fig. \ref{fig:phantom}, are used for testing the proposed image reconstruction algorithm.
The ICA inserts vary in size and concentration; the diameters are 2, 3, 5, and 8 mm and the Iodine
concentrations are 1, 2, 3, and 5 mg/mL. 

\section{Results}
\label{sec:results}

For the present implementation of directional TV minimization, the weighting parameter $\alpha$ is set to 1.75,
favoring the in-plane differentiation in the objective function. The $\ell_1$ penalty parameter $\beta$ is set
to 0.1, and the data error constraint $\epsilon$ is 0.01 in terms of root mean square error.
The imaging volume is discretized on a coarse
grid of size $188 \times 417 \times 175$ with cubic voxels 680 microns wide.

Raw slice images from applying
directional TV minimization to the DBT phantom data are shown in the top panels of Fig.~\ref{fig:imageProc},
and the low spatial-frequency artifacts are clearly visible.
The slice shown cuts through all of the ICA objects.
These objects are easily seen in the 49 kV images, and not the 30kV images, because the K-edge of Iodine is 33.2 keV.
The spatial dependence of the artifact is different
for the low and high kV images, and therefore direct image subtraction of the raw images will not yield
good isolation of the ICA distribution. For performing the post-processing, we employed an eighth degree 2D
polynomial expansion (45 terms), and resulting images are shown in the bottom row of Fig.~\ref{fig:imageProc},
where it is seen that the low spatial-frequency artifacts are greatly reduced.

The results of weighted subtraction for directional TV minimization
filtered back-projection (FBP) are shown in Fig.~\ref{fig:inplane},
where it is seen in the bottom row that it
effectively suppresses the background variation and isolates the ICA objects in the image.

\begin{figure*}[!t]
\centering
\includegraphics[width=0.9\linewidth]{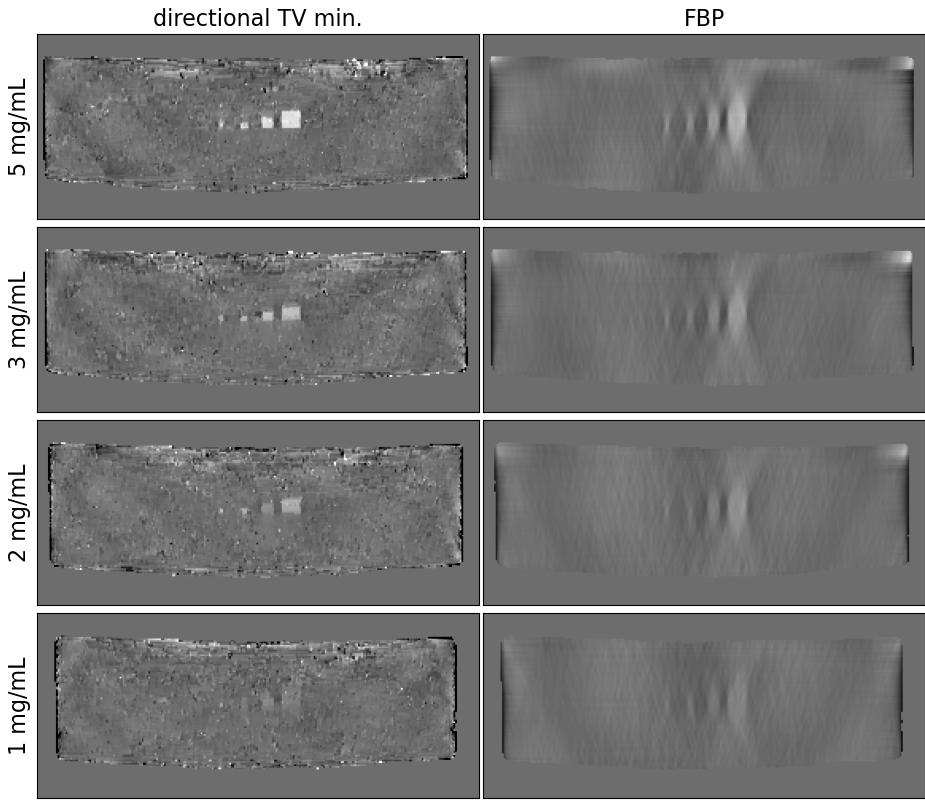}
\caption{Transverse plane images cutting through the groups of ICA objects of the same concentration with
directional TV minimization and FBP images on the left and right, respectively. Concentration of the ICA
decreases going from top to bottom. All gray scales are matched for quantitative comparison and
the images are masked to the object support.}
\label{fig:transverseplane}
\end{figure*}
Transverse slices, where $y$ is horizontal and $z$ is vertical, for the weighted subtraction 
of directional TV minimization and FBP images
are displayed in Fig.~\ref{fig:transverseplane}. While the FBP images show the usual depth blurring
of the ICA objects, it is clear that directional TV minimization can effectively mitigate this artifact.
On the other hand, the directional TV difference images have slightly stronger background clutter
due to imperfect
cancellation of the background glandular structures, which also interferes with the imaging of
the 1mg/mL ICA inserts. This imperfect cancellation is likely due to the non-linearity of the
sparsity regularization.

\section{Conclusion}
In this preliminary study on image reconstruction by directional TV minimization for CE-DBT,
the physical phantom results show promise in achieving an accurate ICA distribution estimate
with nearly isotropic spatial resolution; it appears that the depth blurring artifact common
to DBT imaging is well controlled with directional TV minimization.  Because the dual-energy
processing, in this work, is image-based the gray values in the ICA images are not quantitive
and cannot be directly tied to a concentration value. Future work will focus on spectral
calibration of the DBT transmission data so that both one-step and
two-step \cite{chen2022accurate} quantitative DE-DBT
image reconstruction can be developed. Adapting directional TV minimization to a quantitive DE-DBT
is expected to greatly reduce background clutter in the ICA images from glandular tissue, and this
should also provide accurate ICA concentrations.

\section*{Acknowledgment}
This research was supported in part by the National Institute of Biomedical Imaging and Bioengineering
of the National Institutes of Health under award numbers R01EB023968 and R21CA263660.
The CEDBT system and clinical research was supported financially by Siemens Medical Solutions, USA.



\begin{thebibliography}{1}
\providecommand{\url}[1]{#1}
\csname url@samestyle\endcsname
\providecommand{\newblock}{\relax}
\providecommand{\bibinfo}[2]{#2}
\providecommand{\BIBentrySTDinterwordspacing}{\spaceskip=0pt\relax}
\providecommand{\BIBentryALTinterwordstretchfactor}{4}
\providecommand{\BIBentryALTinterwordspacing}{\spaceskip=\fontdimen2\font plus
\BIBentryALTinterwordstretchfactor\fontdimen3\font minus
  \fontdimen4\font\relax}
\providecommand{\BIBforeignlanguage}[2]{{%
\expandafter\ifx\csname l@#1\endcsname\relax
\typeout{** WARNING: IEEEtran.bst: No hyphenation pattern has been}%
\typeout{** loaded for the language `#1'. Using the pattern for}%
\typeout{** the default language instead.}%
\else
\language=\csname l@#1\endcsname
\fi
#2}}
\providecommand{\BIBdecl}{\relax}
\BIBdecl

\bibitem{sidky2008image}
E.~Y. Sidky and X.~Pan, ``Image reconstruction in circular cone-beam computed
  tomography by constrained, total-variation minimization,'' \emph{Phys. Med.
  Biol.}, vol.~53, pp. 4777--4808, 2008.

\bibitem{zhang2021directional}
Z.~Zhang, B.~Chen, D.~Xia, E.~Y. Sidky, and X.~Pan, ``Directional-{TV}
  algorithm for image reconstruction from limited-angular-range data,''
  \emph{Medical Image Analysis}, vol.~70, p. 102030, 2021.

\bibitem{leo2023}
L.~Y. Zhang, E.~Y. Sidky, J.~P. Phillips, Z.~Zhang, B.~Chen, D.~Xia, and
  X.~Pan, ``Limited scanning arc image reconstruction with weighted anisotropic
  {TV}, minimization,'' in \emph{17th International Meeting on Fully 3D Image
  Reconstruction in Radiology and Nuclear Medicine}, 2023, pp. 491--494.

\bibitem{huang2019comparison}
H.~Huang, D.~A. Scaduto, C.~Liu, J.~Yang, C.~Zhu, K.~Rinaldi, J.~Eisenberg,
  J.~Liu, M.~Hoernig, J.~Wicklein, S.~Vogt, T.~Mertelmeier, P.~R. Fisher, and
  W.~Zhao, ``Comparison of contrast-enhanced digital mammography and
  contrast-enhanced digital breast tomosynthesis for lesion assessment,''
  \emph{J. Med. Imaging}, vol.~6, p. 031407, 2019.

\bibitem{chambolle2011first}
A.~Chambolle and T.~Pock, ``A first-order primal-dual algorithm for convex
  problems with applications to imaging,'' \emph{J. Math. Imag. Vis.}, vol.~40,
  pp. 120--145, 2011.

\bibitem{chen2022accurate}
B.~Chen, Z.~Zhang, D.~Xia, E.~Y. Sidky, T.~Gilat-Schmidt, and X.~Pan,
  ``Accurate image reconstruction in dual-energy {CT} with
  limited-angular-range data using a two-step method,'' \emph{Bioengineering},
  vol.~9, no.~12, p. 775, 2022.

\end{thebibliography}


\end{document}